\documentclass{elsarticle}
\usepackage{url}
\usepackage{amsmath,amssymb,amsfonts}
\usepackage{algorithm}  
\usepackage{mathrsfs}
\usepackage{algpseudocode}  
\usepackage{amsmath}

\usepackage{graphicx}
\usepackage{textcomp}
\usepackage{color,xcolor}
\usepackage{subfig}
\usepackage{tcolorbox}
\usepackage{multicol}
\usepackage{multirow}
\def\BibTeX{{\rm B\kern-.05em{\sc i\kern-.025em b}\kern-.08em
    T\kern-.1667em\lower.7ex\hbox{E}\kern-.125emX}}

\biboptions{authoryear}

\begin{document}
\begin{frontmatter}
% \begin{titlepage}
% \begin{center}
% \vspace*{1cm}

% \textbf{An unsupervised feature selection approach for static analysis warning identification}

% % \vspace{1.5cm}

% % Author names and affiliations
% % Xiuting Ge$^{a}$ (dg20320002@smail.nju.edu.cn), Yifan Huang$^a$ (second.author@mail.com), Mingshuang Qing$^b$ (), Chunrong Fang$^a$ (last.author@mail.com) \\

% % \hspace{10pt}

% % \begin{flushleft}
% % \small  
% % $^a$ The State Key Laboratory for Novel Software Technology, Nanjing University, China \\
% % $^b$ Department of computer science and technology, Southwest University of Science and Technology, China 
% % % $^c$ Full address of last author, including the country name

% % % \begin{comment}
% % % Clearly indicate who will handle correspondence at all stages of refereeing and publication, also post-publication. Ensure that phone numbers (with country and area code) are provided in addition to the e-mail address and the complete postal address. Contact details must be kept up to date by the corresponding author.
% % % \end{comment}

% % \vspace{1cm}
% % \textbf{Corresponding Author:} \\
% % Chunrong Fang \\
% % The State Key Laboratory for Novel Software Technology, Nanjing University, China  \\
% % Tel: (555) 555-1234 \\
% % Email: last.author@mail.com

% % \end{flushleft}        
% \end{center}
% \end{titlepage}

\title{Static Code Analyzer Recommendation via Preference Mining}

\author[nju]{Xiuting Ge}
\ead{dg20320002@smail.nju.edu.cn}

\author[nju]{Chunrong Fang \corref{cor}}
\ead{fangchunrong@nju.edu.cn}

\author[nju]{Xuanye Li}
\ead{522022320077@smail.nju.edu.cn}

\author[nju]{Ye Shang}
\ead{522023320132@smail.nju.edu.cn}

\author[nju]{Mengyao Zhang}
\ead{221250042@smail.nju.edu.cn}

\author[swust]{Ya Pan}
\ead{panya@swust.edu.cn}

\address[nju]{The State Key Laboratory for Novel Software Technology, Nanjing University}
\address[swust]{Department of computer science and technology, Southwest University of Science and Technology}
\cortext[cor]{Corresponding author}

\begin{abstract}
% Static Code Analyzers (SCAs) have been widely applied to detect defects in projects under test. However, SCAs with various static analysis techniques suffer from different levels of false positives and false negatives, thereby yielding the varying effectiveness in SCAs. 
% To detect more defects in a project, it is a possible way to use as many available SCAs for scanning this project as possible. 
% Yet, it is very impractical to simply invoke all available SCAs for a given project in real scenarios.  
% SCA recommendation ------
Static Code Analyzers (SCAs) have played a critical role in software quality assurance. However, SCAs with various static analysis techniques suffer from different levels of false positives and false negatives, thereby yielding the varying performance in SCAs. 
To detect more defects in a given project, it is a possible way to use more available SCAs for scanning this project. 
Due to producing unacceptable costs and overpowering warnings, invoking all available SCAs for a given project is impractical in real scenarios. 
To address the above problem, we are the first to propose a practical SCA recommendation approach via preference mining, which aims to select the most effective SCA for a given project. 
Specifically, our approach performs the SCA effectiveness evaluation to obtain the correspondingly optimal SCAs on projects under test.
Subsequently, our approach performs the SCA preference mining via the project characteristics, thereby analyzing the intrinsic relation between projects under test and the correspondingly optimal SCAs. 
Finally, our approach constructs the SCA recommendation model based on the evaluation data and the associated analysis findings.
We conduct the experimental evaluation on three popular SCAs as well as 213 open-source and large-scale projects. The results present that our constructed SCA recommendation model outperforms four typical baselines by 2 $\sim$ 11 times.
\end{abstract}

\end{frontmatter}

% \vspace{5em}

\section{Introduction} \label{intro} 
% Software defect detection in the project under test as early and many as possible can help reduce software maintenance costs and improve the software quality \citep{kumar2013economics}. 
% Software defect detection, especially considering the situation of rapid delivery and restricted resources, is a challenging task \cite{song2018comprehensive}. 

% The automatic defect detection 
% In the rapidly evolving software lifecyle, 

% functionality correctness, 
The automatic defect detection can help reduce software maintenance costs and improve the software quality \citep{kumar2013economics, tdscme}.
Static Code Analyzers (SCAs) can automatically detect defects without actually executing the program \citep{ayewah2008using}. 
Due to the high code coverage and potential defect detection capability \citep{sca3}, SCAs have already played a key role in software defect detection of open-source and even commercial software projects \citep{imtiaz2019developers, wagner2008evaluation, sonarqubeuse}. 

SCAs, equipped with various static analysis techniques (e.g., data/control flow analysis), have been designed for software projects with different programming languages (e.g., Java and C/C++) and different categories of defects (e.g., performance issues and vulnerabilities) \citep{toolusability}. 
However, these SCAs yield different degrees of False Positives (FPs) and False Negatives (FNs) because of many possible reasons such as the inherent problem of SCAs (i.e., a trade-off between soundness and completeness during the static analysis) \citep{rice} and the implementation bugs of SCAs \citep{limit1, limit2}. 
Consequently, these SCAs depict different levels of effectiveness on the same project. 
To detect more defects in a project under test, it is a possible way to use more available SCAs for scanning this project. Yet, the combination of multiple SCAs could simultaneously produce an overwhelming number of warnings, thereby greatly increasing the manual inspection effort of developers \citep{ge2023machine}.
Also, invoking all available SCAs for a project under test can cause unacceptable costs. 
It is observed that most open-source projects often use only one SCA in open-source projects for software defect detection in practice \citep{beller2016analyzing}.
It indicates that the above way is very impractical in real scenarios, especially for quickly-released projects. 

To alleviate the above problem, an intuitive and potential way is to select the most appropriate SCA for a specific project from a set of available SCAs.
Currently, many studies \citep{rutar2004comparison, sca3, precision5, wagner2005comparing, precision10, liuhan, wangshuai, likaixuan, issta22, toolfn, recall1} have offered clues to perform the SCA recommendation for a specific project by evaluating the effectiveness of SCAs on the defect dataset collected from a set of selected projects. 
However, these studies face the following limitations.
% 耗时耗力，不切实际
First, since the oracle about all ground-truth defects of a given project is unknown \citep{jahangirova2017oracle}, it is difficult to acquire the holistic effectiveness of an SCA on a given project.  
% The SCA often reports a large number of warnings \citep{ge2023machine}, which make the effectiveness of this SCA on a given project tedious.   
Also, invoking all available SCAs for a given project can cause overpowering warnings and unacceptable costs
It suggests that the SCA effectiveness evaluation way in these studies is impractical for the SCA recommendation. 
% 这些研究结果可能不具备泛化性
Second, these studies determine a superior SCA on the defect-oriented dataset rather than the project-oriented dataset.
That is, instead of exploring the defect detection capability of SCAs on a given project, these studies aim to evaluate the holistic effectiveness of SCAs on various types of defects.
It indicates that the findings of these studies may not be generalizable to specific projects. 
% 最优工具可能与项目属性相关
Third, although these studies can provide some hints to recommend an appropriate SCA for a specific project, it is still unclear what makes an SCA work well for this project. 
The effectiveness of SCAs is likely to be project-dependent, which requires an analysis of the project characteristics that impact the effectiveness of these SCAs. 
An SCA performs well in a certain project because it may grasp some particular characteristics of this project. 
Therefore, it is crucial to have a deep understanding of the conditions under which SCA performs well on a specific project.
However, this has been rarely mentioned in the existing studies.
% Such an analysis can help practitioners recommend the most effective SCA for their projects. 

In this paper, we are the first to propose a practical SCA recommendation approach via preference mining. 
Our approach is inspired by the previous instance space analysis studies \citep{sbstinspire1, arpinspire2, arpinspire3} in the search-based software testing and automatic program repair fields. 
These studies are concerned with the objective performance evaluation of different algorithms in the problem instances and the impact of algorithm selection on problem instances. 
These studies further expand Rice's framework \citep{asperti2008intensional} by analyzing why some algorithms might be more or less suited to certain problem instances \citep{arpinspire2}.
By inheriting and extending previous studies, our approach first evaluates the effectiveness of SCAs on massive projects. 
Then, our approach performs the preference mining to deeply analyze why a specific SCA can excel in certain projects. 
Finally, our approach constructs the SCA recommendation model based on the evaluation data and the associated analysis findings.

% To support the candidate SCA effectiveness evaluation, 
Firstly, our approach evaluates the effectiveness of SCAs on massive large-scale and real-world Java projects. 
Instead of the defect-oriented dataset in the existing studies \citep{rutar2004comparison, sca3, precision5, wagner2005comparing, precision10, liuhan, wangshuai, likaixuan, issta22, toolfn, recall1}, our approach performs the effectiveness evaluation of SCAs on the project-oriented dataset by the following operations. 
Specifically, our approach adopts an advanced closed-warning heuristics \citep{dataset11} to each individual project, thereby acquiring reliable labels of warnings reported by SCAs. 
Subsequently, our approach performs the warning alignment to identify identical warnings reported by SCAs, which aims to solve the problem of inconsistent attributes and various quantities of warnings reported by different SCAs.
Finally, our approach fuses FP and FN rates for the SCA effectiveness evaluation, thereby determining correspondingly optimal SCAs for projects under test. 
In particular, to alleviate the oracle problem of all ground-truth defects in a specific project \citep{jahangirova2017oracle}, our approach approximately combines distinct defects detected by SCAs as the ground-truth defects of this project. 

Secondly, our approach performs the SCA preference mining to analyze the relation between the effectiveness of SCAs and projects under test. 
In our approach, the SCA preference is a set of project characteristics where an SCA excels in projects under test.
Specifically, our approach first extracts different granularities (i.e., package, file, class, and method) of characteristics from projects under test.
Subsequently, our approach grasps the most representative project characteristics via the feature selection to link the relation between projects under test and the correspondingly optimal SCAs.
To more clearly understand this relation, our approach performs the feature visualization to witness the distribution of optimal SCAs in projects under test. 

Thirdly, our approach considers the SCA recommendation as a multi-label classification task, thereby constructing an SCA recommendation model based on the correspondingly optimal SCAs of projects under test and the most representative project characteristics.
Given a targeted project, such a model is used to determine the most appropriate SCA for this project from available SCAs.

% That is, the SCA preference in our approach refers to a set of project characteristics where an SCA excels in projects under test.

% our approach considers the SCA recommendation as a multi-label classification task. That is, our approach constructs an SCA recommendation model based on the above history data and uses this model to recommend the most appropriate SCAs for targeted projects.
% To the best of our knowledge, our approach is the first to perform a practical SCA recommendation for 

We conduct the experimental evaluation on three popular SCAs (i.e., SpotBugs\footnote{https://spotbugs.github.io/}, PMD\footnote{https://pmd.github.io/}, and SonarQube\footnote{https://www.sonarsource.com/}) with syntactic and semantic techniques as well as 213 large-scale and open-source Java projects with various domains.
Overall, the experimental results show that in terms of \emph{F$_{\beta}$}($\beta$ = 1), SonarQube can perform better than SpotBugs and PMD on 213 projects. 
After analyzing the relation between the project characteristics and the optimal SCA, it is observed that when a project has class-level characteristics with greater values, SonarQube tends to perform higher \emph{F$_{\beta}$}($\beta$ = 1).
Further, we compare our constructed SCA recommendation model with four typical baselines (including the random selection). The results depict that our constructed SCA recommendation model can achieve optimal performance. In particular, our constructed SCA recommendation model is about 2 $\sim$ 11 times better than four typical baselines in terms of P$_{micro}$.

% P$_{micro}$ with 0.63, which is *** better than four baselines in terms of respectively. 

% Such an analysis helps reveal the pros and cons of SCAs in the defect detection capability of different projects.

The main contributions of this paper are as follows.
\begin{itemize}
    \item \textbf{Preference analysis.} Based on the extracted project characteristics, we perform an SCA preference mining to deeply analyze the intrinsic relation between the performance of SCAs and projects under test. 
    
    \item \textbf{Practical approach.} To the best of our knowledge, we are the first to propose an SCA recommendation approach via preference mining. 
    Such an approach can provide the practical SCA recommendation for specific projects.
    
    \item \textbf{Effectiveness evaluation.} By conducting the experimental evaluation on three typical SCAs as well as 213 large-scale and open-source Java projects, the results present that our approach performs more effective SCA recommendation than four typical baselines. 

    \item \textbf{Available artifacts.} We share all associated artifacts (including datasets and scripts) in a public repository \citep{mylink}, which facilitates following and extending our research.
\end{itemize}

The remainder of this paper is organized as follows. Section \ref{back} presents the background and related work. Section \ref{approach} shows the overview of our approach. Section \ref{setup} describes the experiment setup. Section \ref{result} gives the experimental evaluation results and analysis. Section \ref{discussion} discusses implications and threats to the validity. Section \ref{conclusion} concludes this paper.

\section{Background and related work} \label{back} 

% 介绍sca扮演的重要作用，最新的tse论文参考，介绍工具的典型分类
% (e.g., pattern matching, abstract interpretation, or data/control flow analysis)
\subsection{SCAs} \label{sca}
\begin{figure}
    \centering
    \includegraphics[width=1\linewidth]{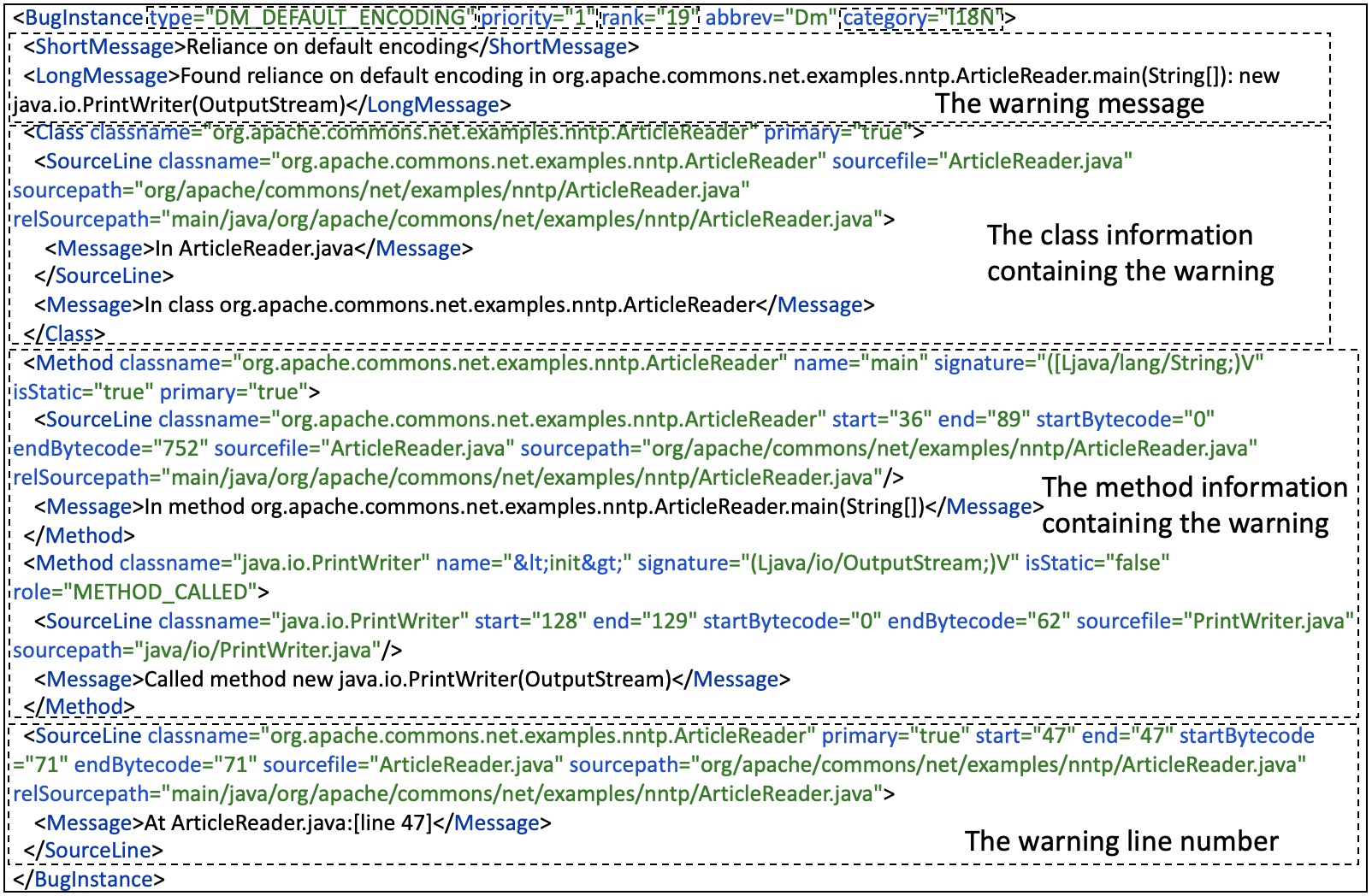}
    \caption{An example of a warning reported by SpotBugs.}
    \label{fig:example}
\end{figure}

SCAs automatically analyze the source code or binary code for software defect detection by using various static analysis techniques \citep{Statictech}.
In essence, these SCAs work through syntactic and semantic analysis \citep{likaixuan}. 
Thus, the existing SCAs can be mainly divided into syntax- and semantics-based SCAs. 
To help developers quickly and easily locate and understand defects in the project under test, warnings reported by SCAs generally consist of the category, severity, message, and location. 
Of these, the warning location is composed of the class and method information containing a warning and the warning line numbers. 
Figure \ref{fig:example} shows a specific warning reported by SpotBugs.

% 介绍SCA实证研究
\subsection{Empirical Studies on SCAs} 
Currently, many studies have conducted the warning quantitative analysis to empirically evaluate the effectiveness of SCAs. 
On the one hand, some studies \citep{rutar2004comparison, sca3, precision5, wagner2005comparing, precision10, liuhan, wangshuai} aim to discern whether there are FPs in the warnings reported by SCAs from the perspective of soundness. 
On the other hand, some studies \citep{likaixuan, liuhan, wangshuai, issta22, toolfn, recall1} are concerned with whether there are FNs in the warnings reported by SCAs from the perspective of completeness. 
The above studies demonstrate the varying effectiveness of different SCAs in software defect detection, which provide some hints for the SCA recommendation. However, these studies evaluate the SCA effectiveness on the defect-oriented dataset rather than the project-oriented dataset. 
Also, these studies rarely explore the relation between SCAs and projects under test.
Consequently, such phenomena could hinder the SCA recommendation for given projects.
% Despite providing some hints for the SCA recommendation, these studies rarely focus on the relation between SCAs and projects under test, which could hinder the effective SCA recommendation for targeted projects.

% 为什么没有考虑多个工具组合进去
% 介绍inspire的几个工作
\subsection{Related Work} 
With the emergence of different SCAs and their varying defect detection capabilities, researchers and practitioners attempt to gather multiple SCAs for a project under test. 
Despite reducing FNs through the complementarity of different SCAs \citep{mapping2}, the combination strategy of multiple SCAs can significantly increase FPs \citep{ge2023machine}. 
Also, few projects in realistic scenarios use multiple SCAs for defect detection simultaneously \citep{beller2016analyzing}. 
Thus, our approach mainly focuses on selecting one optimal SCA for a targeted project. 

% 我们的方法是场景无关的，选择最合适的准则来评估静态分析工具的有效性，目标不同

Some studies propose to recommend the optimal algorithm for a given problem instance \citep{sbstinspire1, arpinspire2, arpinspire3}. For example, Chen et al. \citep{sbstinspire1} optimize the test prioritization by analyzing the test distribution. 
Aldeida et al. \citep{arpinspire2} analyze the relation between projects under test and Automated Program Repair (APR) tools and rely on such relation for APR tool recommendation.
Zhong et al. \citep{arpinspire3} mine the repair patterns from fixed bugs and design a preference-based ensemble approach to recommend an APR tool for a given bug.
However, there are few studies for the SCA recommendation. 
Fortunately, our approach, fundamentally an instance space analysis problem, is similar to the core ideas of the above studies. 
As such, our approach inherits and extends the core ideas of the above studies to recommend the most appropriate SCA for a specific project.

\subsection{Problem Statement}  \label{ps}
Our approach denotes the SCA recommendation as a supervised learning-based multi-classification problem. 
Formally, given a set of projects \emph{P = $\{p_1, ...,\\ p_i, ..., p_n\}$} (\emph{n} is the number of projects) and a set of available SCAs \emph{S = $\{s_1, ..., s_i,\\ ..., s_m\}$} (\emph{m} is the number of available SCAs), 
an optimal SCA \emph{$s_{best}$} can be determined for each project \emph{$p_i$} by evaluating the effectiveness of \emph{S} on \emph{$p_i$}. 
After that, our approach can collect the training, which is represented as \emph{(P, S)}. Of these, \emph{($p_i, s_{best}$)} (\emph{$p_i \in P$}, \emph{$s_{best} \in S$}) is a training sample. 
Second, our approach extracts the significant characteristics from \emph{P} as the feature vector \emph{$F \in \rm{R}^d$}, where \emph{d} is the dimension of $F$. 
Third, our approach learns a decision function \emph{$f: \rm{R}^d \rightarrow S$} to describe the mapping relations between each project \emph{$p_i$} and the correspondingly optimal SCA \emph{$s_{best}$}. 
When given a targeted project \emph{$p_{target}$}, the learned decision function can recommend the most appropriate SCA \emph{$s_{target}$} (\emph{$s_{target} \in S$}) to \emph{$p_{target}$} for software defect detection.

% Subsequently, our approach extracts the characteristics (e.g., complexity) of \emph{$p_i$} as features \emph{$F \in \rm{R}^k = \{f_1, ..., f_i, ..., f_k\}$} (\emph{k} is the number of features in \emph{$p_i$}). 

% given a set of training data \emph{$D = \{(x_1, y_1), (x_2, y_2), ..., (x_n, y_n)\}$}, \emph{$x_i \in \rm{R}^d$} is the feature vector that is extracted from the project characteristics
% Next, our approach 

\section{Approach} \label{approach}

\begin{figure*}[htbp]
\centerline{\includegraphics[width=1\textwidth]{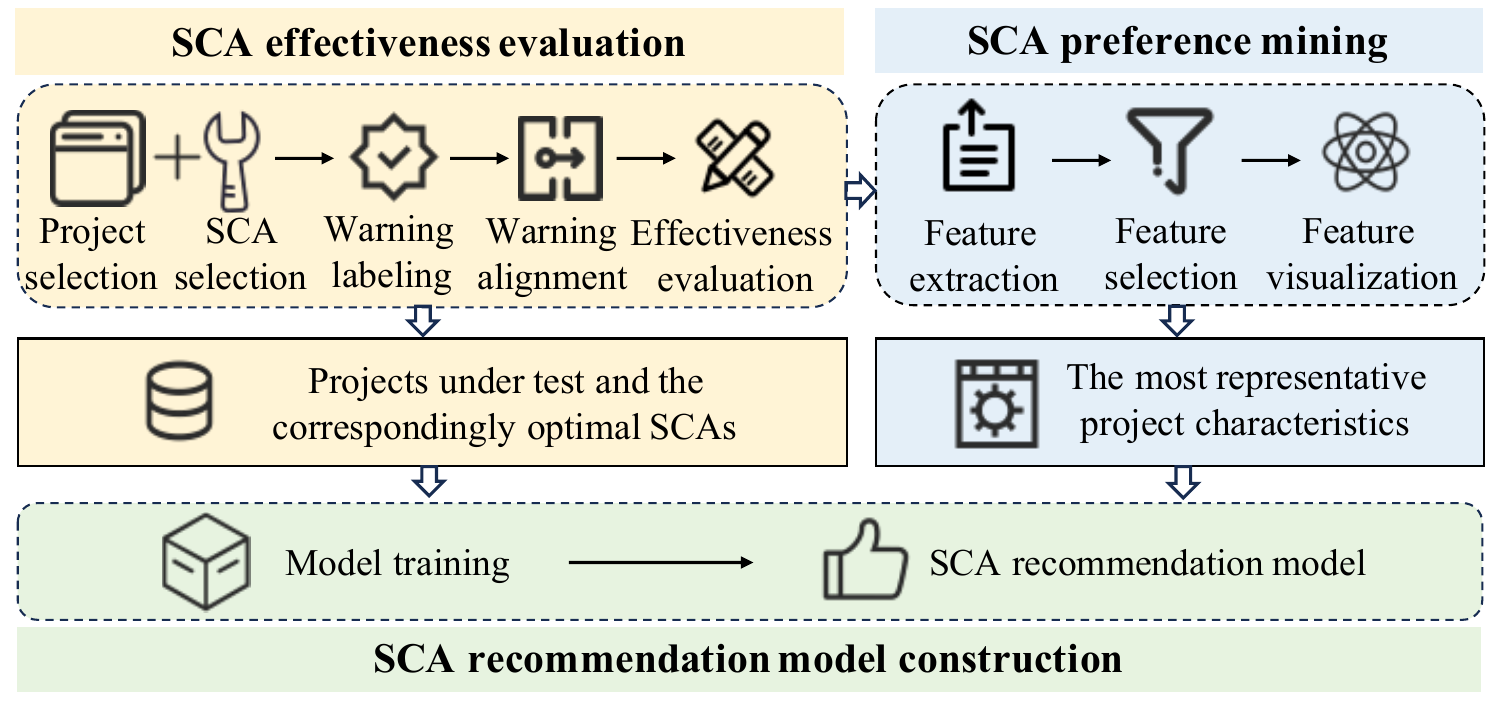}}
\caption{Overview of our approach.}
\label{overview}
\end{figure*}

% Based on Section \ref{ps}, our approach mainly involves SCA performance evaluation, SCA preference mining, and SCA recommendation model construction. 
As shown in Figure \ref{overview}, our approach involves three stages, including \emph{SCA performance evaluation}, \emph{SCA preference mining}, and \emph{SCA recommendation model construction}.
\emph{SCA performance evaluation} empirically studies the effectiveness of SCAs in various projects, thereby searching an optimal SCA for each project. 
\emph{SCA preference mining} performs the feature extraction, selection, and visualization. Such an operation aims to identify the most representative features between projects under test and the correspondingly optimal SCAs.
Based on the results of the first and second stages, \emph{SCA recommendation model construction} leverages Machine Learning (ML) techniques to construct the SCA recommendation model. Such a model is used to determine the most appropriate SCA for a targeted project.
Based on the above overview of our approach, we further show the applications of our approach.

\subsection{SCA Performance Evaluation} \label{evaluation} 
Our approach first selects the representative projects and typical SCAs by designing the rigorous selection criteria.
When our approach uses a determined SCA to scan a selected project, a series of warnings can be obtained. Subsequently, our approach performs reliable warning labeling by refining an advanced closed-warning heuristic \citep{dataset11}. 
Since different SCAs have different warning output formats, our approach performs the accurate warning alignment by constructing a unified mapping relation. 
After that, our approach empirically performs the detailed effectiveness evaluation for SCAs. 
Finally, our approach can search the correspondingly optimal SCAs for  projects under test. 

\subsubsection{Project selection} \label{projectselection}
Our approach considers 730 curated Java projects of Github repository\footnote{https://github.com/.} in the work of Liu et al. \citep{miningha} as the start point for our project selection. 
Subsequently, our approach performs the project selection by rigorously designing the following criteria. 
First, each project has various domains (e.g., database, Java utility, and network application), which ensures that the selected projects have enough representativeness. 
Second, each project has more than 500 stars, which aims to ensure that the selected projects are notable. Third, each project has at least 500 commits from the project creation time to the warning collection time, which ensures that the selected projects have sufficient maturity. Fourth, each project can be compatible with Java8, which ensures that the selected projects can be scanned by SCAs. Fifth, each project has at least two compilable releases (a.k.a., tags), which ensures that SCAs can run on the releases successfully and the warning labeling can be supported. Sixth, each project is developed collaboratively by multiple developers, which ensures that each project can fully reflect the real-world scenario. 
To the end, our approach selects 213 Java projects from 730 curated projects as the research target.
% 由于空间限制，本文将选择的项目展示在了开源仓库中

\subsubsection{SCA selection} \label{scaselection}
Our approach rigorously designs the following criteria for SCA selection. First, each SCA can support to scan Java projects. Second, each SCA is open-source. Although commercial SCAs are prevalent in the industry, they often entail substantial costs.
Third, each SCA is under maintenance, which ensures that the selected SCAs have sufficient recognition.
Fourth, the selected SCAs span various static analysis techniques, which are mainly classified into syntactic (pattern matching) and semantic (data flow and control flow) techniques in Section \ref{sca}. It ensures that the selected SCAs are representative enough. 
Based on the above criteria, our approach selects three typical and commonly used SCAs, including the syntax-based SpotBugs, the syntax-based PMD, and the semantics-based SonarQube.
% Fifth, the selected SCAs can be integrated for ---需要写持续集成这个优点不

% Table \ref{tab:scaetail} shows the selected SCAs with  techniques and versions.

% \begin{table}
%     \centering
%     \caption{Details about the selected SCAs.}
%     \label{tab:scaetail}
%     \begin{tabular}{|c|c|c|}
%     \hline
%        \textbf{Tool}  & \textbf{Technique} & \textbf{Version}\\ \hline \hline
%         SpotBugs & Syntactic & v4.0\\ \hline
%         PMD & Syntactic & v7.1.0\\ \hline
%         SonarQube &  Semantic & v7.8.0\\ \hline
%     \end{tabular}
% \end{table}

\subsubsection{Warning labeling} \label{labeling}
After selecting projects and SCAs, our approach uses each SCA to scan each project respectively. 
Each project has many releases, and the release determination can affect the reported warning quantity and the labeled warning quality. 
Given a project with a set of releases, our approach first determines a latest and compilable release \emph{r$_{latest}$}, whose time is closest to the warning collection time in this paper (i.e., 2023/12/20). 
Subsequently, our approach selects an additional and compilable release \emph{r$_{addtional}$}, which comes before \emph{r$_{latest}$}. 
After using an SCA to separately scan \emph{r$_{addtional}$} and \emph{r$_{latest}$} in a project, our approach finally refines an advanced closed-warning heuristic \citep{dataset11} to track the warning evolution, thereby relying on warnings in \emph{r$_{latest}$} to label actionable and unactionable warnings in \emph{r$_{latest}$}. 
In particular, the time interval between \emph{r$_{latest}$} and \emph{r$_{addtional}$} is about 90 days. Such a time interval is generally considered as a software iteration cycle \citep{alshayeb2003empirical}, which can help identify enough actionable warnings in \emph{r$_{addtional}$}. 
If such a time interval is far more than 90 days, the source code between \emph{r$_{latest}$} and \emph{r$_{addtional}$} could undergo significant changes, which may cause the closed-warning heuristic to label massive unknown warnings. 
If such a time interval is far less than 90 days, the source code between \emph{r$_{latest}$} and \emph{r$_{addtional}$} could be very similar, which may cause the closed-warning heuristic to label few actionable warnings. 
The above phenomena (i.e., too many unknown warnings and too few actionable warnings) could fail to fully reflect the effectiveness differences of SCAs in projects. 
% may affect SCA performance evaluation, thereby .

The core idea behind such a heuristic is that (1) if a warning in \emph{r$_{addtional}$} disappears in \emph{r$_{latest}$}, this warning is labeled to be actionable.
If a warning in \emph{r$_{addtional}$} is present in \emph{r$_{latest}$}, this warning is labeled to be unactionable. 
If the class/method, where a warning is in \emph{r$_{addtional}$}, is deleted in \emph{r$_{latest}$}, this warning is labeled to be unknown. 
In such a heuristic, it is critical to judge whether two warnings are identical between two releases. 
The existing studies \citep{heu1, golden, modelbuilding} perform the pair-wise matching via the exact matching, i.e., judge two warnings by comparing the warning type and the warning location.
However, the exact matching misses the pair-wise warnings when the class/method containing the warning in the warning location is renamed/refactored or undergoes drastic source code changes. 

To mitigate the above problem, our approach refines such a heuristic by incorporating three-stage warning matching strategy \citep{miningha}.
Given a warning \emph{w$_a$} in \emph{r$_{addtional}$} and a warning \emph{w$_b$} in \emph{r$_{latest}$}, the specific process of such a strategy is shown as follows. 
(1) \emph{The location-based matching} computes the offset of warnings in code change diffs. If \emph{w$_a$} and \emph{w$_b$} are in the offset of code change diffs and the offset is not more than 3, \emph{w$_a$} and \emph{w$_b$} are considered to be identical. 
The location-based matching way can tolerate a slight code location change related to the warning line numbers while requiring the same class path and method path in \emph{w$_a$} and \emph{w$_b$}.
When the location-based matching way fails, (2) the snippet-based matching way is used to compare the warning text strings. If \emph{w$_a$} and \emph{w$_b$} have the same code strings, \emph{w$_a$} and \emph{w$_b$} are considered to be identical. 
The snippet-based matching way can be resilient to a significant code location change related to the warning line numbers while requiring the same class path. 
When the snippet-based matching way fails, (3) the hash-based matching way is used to compute the hash values of adjacent 100 tokens around the warning. If \emph{w$_a$} and \emph{w$_b$} have the same hash value, \emph{w$_a$} and \emph{w$_b$} are considered to be identical. 
The hash-based matching way can handle class/method renaming and source code refactoring to some extent.
Based on such a refined heuristic, our approach can label actionable, unactionable, and unknown warnings in \emph{r$_{addtional}$}. 
It is noted that our approach only focuses on actionable and unactionable warnings for SCA performance evaluation. 
Ultimately, our approach obtains 639 reports (3 SCAs * 213 projects), where each report has different numbers of labeled warnings. 

\begin{figure}
    \centering
    \includegraphics[scale=0.5]{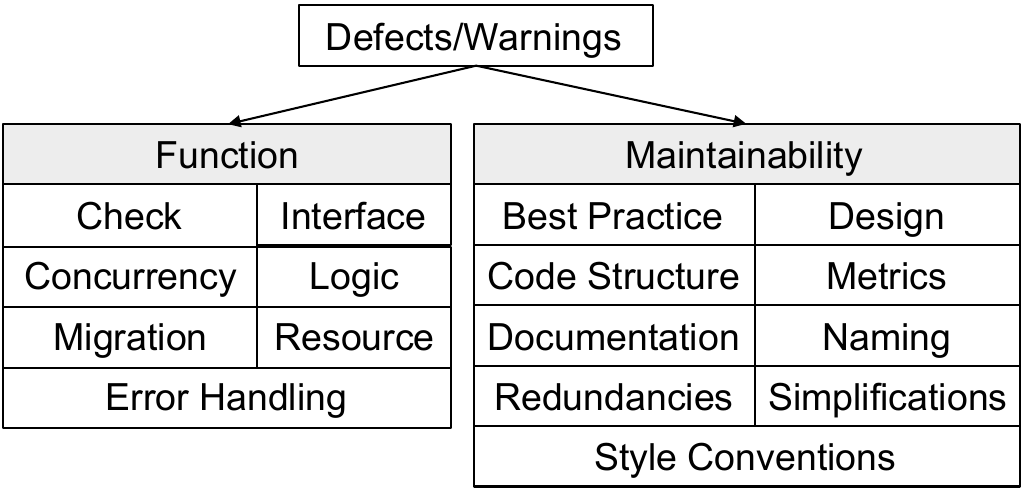}
    \caption{General defect classification}
    \label{fig:gdc}
\end{figure}

% If directly 
% As such, it is impractical to 
% directly performing the warning alignment by exactly matching attributes of warnings reported by three SCAs 

% Second, some warning attributes are incomplete among three SCAs. For example, SpotBugs has the method information containing a warning. By contrast, PMD and SonarQube have no  method information related to a warning. 
% Third, the warning attributes with the same meaning are named differently among SCAs. For example, the rank in SpotBugs (i.e., 1$\sim$20) and the severity in SonarQube (i.e., BLOCKER, CRITICAL, MAJOR, and MINOR) denote whether a warning needs to be inspected first. 
% Consequently, the above phenomena may misjudge identical warnings reported by different SCAs, thereby affecting the SCA performance evaluation.

% By contrast, the severity in SonarQube is denoted as the string (i.e., BLOCKER, CRITICAL, MAJOR, and MINOR). However, both the rank and severity are

% affect the identical warning judgement when using different
% misjudge two identical warnings reported by different SCAs as two different ones.

\subsubsection{Warning alignment} \label{alignment}
% Our approach performs the warning alignment to discern identical warnings reported by three SCAs in the same project. 
The warnings among three SCAs display different attributes, especially for the warning type.  
For example, in the file ``net.finmath.time.businessdaycalendar.\\BusinessdayCalendar.java'' of the ``finmath-lin'' project, SpotBugs reports a warning with the type ``DM\_BOXED\_PRIMITIVE\_FOR\_PARSING'' in line 139. 
Similar, SonarQube also report a warning with the type ``CODE\_SMELL'' in line 139. 
Although the types of these two warnings are different, the issue pinpointed by these two warnings is the same. 
It indicates that these two warnings should be considered to be identical ones. 
If the warning alignment is directly performed by exactly matching the original attributes of warnings reported by three SCAs, the inconsistent warning types may miss many identical warnings reported by three SCAs, thereby affecting the SCA performance evaluation. 
To address the above problem, our approach performs the warning alignment to discern identical warnings reported by three SCAs in the same project. The detailed process of warning alignment is shown as follows.

First, our approach constructs a unified mapping relation to address the inconsistent warning types. 
Specifically, our approach relies on General Defect Classification (GDC) \citep{mapping1, mapping2, mapping3} to align the inconsistent warning types among SCAs.
As shown in Figure \ref{fig:gdc}, GDC involves two high-level and 16 fine-grained defect categories. Our approach extracts all original warning types in each SCA and classifies each original warning type into fine-grained defect categories via manual inspection. Such a process is separately conducted by two authors in this paper. 
If the classification result of the two authors is identical, our approach records the classification results. Otherwise, the two authors make a discussion and arrive at a consistent classification result. 

Second, our approach uniformly defines a warning as \emph{$w$ = (newType, classInfo, startLine, endLine, warningLabel)}. 
Of these, the original warning types in three SCAs are re-classified into \emph{newType}. 
\emph{classInfo} is the class information containing \emph{w}. 
\emph{startLine} and \emph{endLine} determine the warning line numbers. 
\emph{warningLabel} contains actionable and unactionable ones obtained in Section \ref{labeling}. 
It is noted that our approach does not consider the method information containing the warning and the warning severity/priority/rank into \emph{w}. It is because the current warning attributes in \emph{w} have clearly described an issue reported by a warning and uniquely determined a warning location. 
Despite enriching the warning information, the method information containing the warning and the warning severity/priority/rank are trivial for SCA effectiveness evaluation.  
Based on the definition of \emph{w}, our approach can convert all warnings in 639 reports of Section \ref{labeling} into the same format.

Third, our approach judges whether warnings reported by three SCAs in a project are identical. 
Given any three warnings reported by all SCAs in the same project, if (1) the \emph{newType}, \emph{classInfo}, and \emph{warningLabel} are the same; (2) the offset of \emph{startLine} and \emph{endLine} is within 3 respectively; and (3) the scope from \emph{startLine} and \emph{endLine} has the overlap. These three warnings are considered to be identical. 
Noted, the offset setting in the second condition can be resilient to the fact that SCAs have slight differences in locating the code lines of the same warning.  
%===
If the above three conditions except \emph{warningLabel} are satisfied among three warnings, our approach follows the voting mechanism to uniformly assign the same warning label for these three warnings. 
%====
If the above three conditions except \emph{warningLabel} in any two warnings are satisfied, these two warnings are discarded. 
This is because the inconsistent warning labels yield biased SCA effectiveness evaluation. 
%====
If these three warnings are distinct each other, they are retained. Such an operation facilitates identifying the preferences of different SCAs. 
%
% The above warning matching operation is performed for each three SCA reports in a project. In all, such an operation is executed 213 times.   

% As for warning 
% To the end, as for three reports 
% the identical actionable and unactionable warnings in 613 reports of Section \label{labeling} 

% If all attributes except \emph{warningLabel} among warnings are identical, our approach follows the voting mechanism to assign a new warning label for the three warnings.  
%
% In particular, if warnings reported by any two SCAs have the same \emph{newType}, \emph{classInfo}, \emph{methodInfo}, and \emph{warningLineNum} but have different \emph{warningLabel}, these warnings are discarded.
% It is because the warning label is ambiguous, thereby causing the biased SCA performance evaluation. 
% If warnings are only reported by any an SCA, these warnings are retained, which helps identify the preference of different SCAs. 
% Otherwise, \emph{w$_a$} and \emph{w$_b$} are considered to be different.

\subsubsection{Effectiveness evaluation} \label{effecteva} 
% 准则选择
To perform the effectiveness evaluation for SCAs, our approach selects \emph{F$_{\beta}$} as a measurement. 
Equation (\ref{f}) shows the calculation details for \emph{F$_{\beta}$}. 
Of these, True Positive (\emph{TP}) and False Negative (\emph{FN}) describe the number of actionable warnings that are correctly and incorrectly identified by an SCA respectively. False Positive (\emph{FP}) describes the number of unactionable warnings that are incorrectly identified by an SCA. Precision (\emph{P}) is the ratio of reported actionable warnings in all warnings reported by an SCA. Recall (\emph{R}) is the ratio of reported actionable warnings in all actionable warnings of a project.
In particular, $\beta$ is a hyperparameter. If $\beta$ $\textgreater$ 1, \emph{F$_{\beta}$} focuses more on \emph{R}. If $\beta$ = 1, \emph{F$_{\beta}$} gives equal weight to \emph{R} and \emph{P}. If $\beta$ $\textless$ 1, \emph{F$_{\beta}$} focuses more on \emph{P}.
The reason for selecting \emph{F$_{\beta}$} as a measurement is that (1) \emph{P} and \emph{R} are the most commonly used in the previous SCA effectiveness evaluation studies \citep{toolfn, liuhan, likaixuan, lipp2022empirical}, (2) \emph{F$_{\beta}$} can juggle the precision and recall, and (3) $\beta$ in \emph{F$_{\beta}$} can support smoothly adjusting developer requirements for SCA effectiveness evaluation.

{
\small
\begin{equation}
    F_{\beta} = \frac{(1+{\beta}^2)*P*R}{({\beta}^2*P)+R}, 
    \\
    P = \frac{TP}{TP + FP}, \\
     R = \frac{TP}{TP + FN}
    \label{f}
\end{equation}
}

% \begin{equation}
%     F_{\beta} = \frac{(1 + {\beta}^2) * P * R}{({\beta}^2 * P) + R}
%     \label{f}
% \end{equation}

% \begin{equation}
%     P = \frac{TP}{TP + FP}
%     \label{p}
% \end{equation}

% \begin{equation}
%     R = \frac{TP}{TP + FN}
%     \label{r}
% \end{equation}

% 如何确定tp，fp，Fn 
Next, our approach calculates \emph{F$_{\beta}$} for each project and each SCA. 
Given a set of labeled and aligned warnings \emph{W} reported by using an SCA for a project, \emph{TP} and \emph{FP} is the number of all labeled actionable and unactionable warnings in \emph{W}. Thus, \emph{P} can calculated via \emph{TP} and \emph{FP}. 
However, \emph{FN} is unknown because all ground-truth defects in a project are undecidable \citep{jahangirova2017oracle}. 
To address this problem, our approach considers the number of all distinct actionable warnings reported by using three SCAs for a project as ground-truth defects in this project (i.e., \emph{TP}+\emph{FN}). 
The rationale behind such an operation is that the combination of multiple SCAs can increase the defect discovery ability \citep{multiple1, mapping2, multiple2, likaixuan}. 
Although such an operation could miss some ground-truth defects, these omitted defects are not also taking \emph{TP}+\emph{FN} of \emph{R} into account, thereby ensuring fairness in the process of SCA effectiveness evaluation. 

After obtaining the effectiveness of each SCA in each project (i.e., 3 SCAs * 213 projects = 639 \emph{F$_{\beta}$} results), our approach can finally identify the optimal SCA for each project.  
In all, there are 213 samples, which involve 213 projects with the correspondingly optimal SCA.

% After performing the SCA effectiveness evaluation, our approach can identify the optimal SCA for each project. 
% It is noted that our approach does not extract ground-truth defects in a project by automatically tracking the software evolution. 
% It is because the automatic software evolution traceability is easily affected by inaccurate commit messages and tangled commit structure, thereby causing mislabeled defects \citep{yatish2019mining, kim2011dealing}. 
% In contrast, 

% Since the extracted software metric-based features may have irrelevant and redundant ones, 
\subsection{SCA Preference Mining} \label{mining}
To mine the preferences of different SCAs on projects under test, our approach extracts features related to project characteristics. 
Besides, our approach leverages the Recursive Feature Elimination (RFE) algorithm \citep{rfepaper} to select the most representative features from all extracted features. Moreover, our approach performs the feature visualization to enlighten the intrinsic relation between projects under test and the correspondingly optimal SCAs.

% thereby mining preferences of different SCAs on projects under test.

\subsubsection{Feature extraction} \label{extract} 
Our approach extracts project characteristics as features to map the relation between projects under test and the correspondingly optimal SCAs.
The rationale behind such an operation is shown as follows. 
On the one hand, SCA preference mining in our approach is essentially to reveal the software defect detection ability of SCAs. In the software defect detection community, massive existing studies rely on project characteristics (e.g., Line of Code in a class) to detect software defects and have exhibited remarkable performance \citep{sdp1, sdp2}. 
On the other hand, the effectiveness of different SCAs could be related to the selected projects. 
Generally, SCAs on more complex projects perform worse than those on more simple projects \citep{d2a}. 
Fortunately, it has been proved that project characteristics-based features are capable of sufficiently denoting the complexity of a project \citep{heihei, nam2015clami}. 
As such, our approach incorporates project characteristics-based features into mining SCA preferences. 
Though an off-the-shelf Understand tool\footnote{https://scitools.com/.}, our approach extracts 302 features for each project, involving 78 package-related, 80 file-related, 90 class-related, 54 method-related project characteristics. 
% different SCAs employ various syntactic and semantic techniques for automatic software defect detection. 
% SCAs with more delicate static analysis techniques perform better in a project with high complexity.
% Disparities among SCA techniques lead to detecting different software defects. Generally, as for a complex project instances, more delicate and refined SCA techniques perform better at software defect detection \citep{likaixuan, liuhan}.
% 

% https://blog.csdn.net/FontThrone/article/details/79004874/
\subsubsection{Feature selection} \label{select} 
% It is possible that not all extracted features could contribute to isolate ``hard'' and ``easy'' projects.
However, the extracted features may capture redundant project elements or be irrelevant to reveal the SCA preferences.
Thus, it is critical to meticulously select a discriminative feature subset from all extracted features, thereby effectively demystifying the intrinsic relation between projects under test and the correspondingly optimal SCAs.

% for identifying differences ``hard'' and ``easy'' projects,

Our approach uses RFE \citep{rfepaper} for feature selection. 
The core idea behind RFE is relying on the ML model to iteratively remove the weakest features that contribute the least to the model performance, until the desired number of features is achieved. 
The reasons for applying RFE to feature selection are shown as follows. On the one hand, the characteristics of RFE, being good at dealing with a large number of features \citep{rfepaper} and enabling the model to efficiently and easily interpret \citep{jain1997feature}, can be exactly utilized to address the problem of our approach.
On the other hand, compared to the feature subset obtained by other off-the-shelf feature selection algorithms, the feature subset obtained by RFE generally exhibits better explainability and higher model performance in various domains (e.g., actionable warning identification \citep{wang2018is, S18}).

% 交叉验证使用默认cv；模型使用RF
Specifically, RFE first relies on all extracted features to train a Machine Learning (ML) model based on 213 samples obtained in Section \ref{evaluation}. 
Second, the importance of each feature, which is denoted as the coefficient in this ML model, is determined. Based on the importance of features, RFE ranks all extracted features.  
Third, RFE removes features with the least importance from the current feature set. 
Fourth, the above three operations are repeated with the reduced feature set until a pre-specified number of features is left or the further feature removal can result in a significant loss of this ML model performance. 
It is noted that the above RFE needs to manually set the number of the feature subset, which results in the local optimum. 
To alleviate this problem, our approach employs cross-validation with RFE, thereby automatically determining the optimal number of the feature subset. 
At last, our approach selects 100 features from 302 initially extracted features for SCA recommendation model construction.

% RFE is very good at dealing with datasets with a large number of features \citep{rfepaper}. By focusing on the most important features, RFE can improve the model performance \citep{wang2018is, S18} and enable the model to be more efficient and easier to interpret \citep{jain1997feature}.

\subsubsection{Feature visualization} \label{visual}
Based on the most representative feature subset, our approach performs the feature visualization via Principal Component Analysis (PCA) \citep{pca}. 
The core idea of PCA is to transform a large set of variables into smaller ones that still contain the most information in the large set of variables. These smaller variables are called Principal Components (PCs). 
In general, the first PC can explain the maximum amount of variables and the first few PCs can explain almost all variables in the dataset.
Besides, PCs are orthogonal to each other, which ensures that the obtained PCs are linearly independent. 
Moreover, PCA is effective in the dimensionality reduction of dataset, thereby simplifying the complexity in the visual exploratory analysis. 

% , which make the feature visualization more easier. 
Inspired by the characteristics of PCA, our approach incorporates PCA into the feature visualization. 
Specifically, our approach applies PCA to obtain all PCs of 100 selected features across 213 samples in Section \ref{evaluation}. Subsequently, our approach selects the first two PCs for the feature visualization.
Finally, to provide insights for the intrinsic relation between projects under test and the correspondingly optimal SCAs, our approach visualizes (1) the feature footprints to observe how the value distribution of the most representative feature subset in projects under test and (2) the SCA footprints to witness how the distribution of optimal SCAs in projects under test. 

% our approaches visualises the value distribution of the most representative feature subset and the optimal SCAs in the projects. 

% to provide insights for the intrinsic relation between optimal SCAs and  projects, our approach performs the feature visualization from two aspects.
% On the one hand, our approach 
% performs the feature visualization from two aspects.
% visualises the feature footprints to observe how the value distribution of the most representative feature subset in the projects. 
% On the other hand, our approach visualises the SCA footprints to witness how the distribution of optimal SCAs in the project. 

% projects under test and the  optimal SCAs.
\subsection{SCA Recommendation Model Construction} \label{construction}
Based on the data in Section \ref{evaluation} (i.e., projects under test and the correspondingly optimal SCAs) and Section \ref{mining} (i.e., the most representative project characteristics), our approach relies on the multi-label classification algorithm to construct an SCA recommendation model. 
% Specifically, our approach relies on the multi-label classification algorithm for model training based on the data with associated features. Finally, an SCA recommendation model can be obtained. 

 % our approach first uses SMOTE \citep{chawla2002smote} to alleviate the class imbalance problem in the data.

% 定制化的；灵活的；generic;
\subsection{Applications of Our Approach}
Given a specific project, our approach adopts the well-trained model in Section \ref{construction} to recommend an optimal SCA from three candidate SCAs for this project. 

Our approach has two primary advantages. 
On the one hand, our approach is highly customizable. Different developers have different requirements when selecting SCAs for the software defect defection of projects under test. 
Some focus more on soundness by aggressively suppressing unactionable warnings, while others may opt for completeness at the price of more unactionable warnings. 
By adjusting $\beta$ in Section \ref{evaluation}, our approach can flexibly satisfy different requirements of developers on the SCA effectiveness. 
On the other hand, the infrastructure of our approach is generic. 
In addition to three candidate SCAs (i.e., SpotBugs, PMD, SonarQube) studied in this paper, our approach can easily extend the recommendation of other SCAs or the combination of multiple SCAs for specific projects.

\section{Experiment Setup} \label{setup}
\subsection{Research Questions} \label{rq}
Based on the overview of our approach in Section \ref{approach}, we propose the following Research Questions (RQs).

\textbf{RQ1: SCA effectiveness evaluation.} 
\begin{itemize}
    \item How do three candidate SCAs perform?
\end{itemize}

\textbf{RQ2: SCA preference mining.} 
\begin{itemize}
    % 第一个先指出那些特征，然后绘制值分布以解释
    % \item \textbf{RQ2.1} : What features of project are selected for optimal SCAs?
    \item \textbf{RQ2.1}: Which features are the most representative to mine the preferences of candidate SCAs on the projects under test?
    \item \textbf{RQ2.2}: How is the feature distribution between projects under test and the correspondingly optimal SCAs?
    % Which features of project instances can impact the effectiveness of SCAs? 
    % \item \textbf{RQ2.2} : How is the distribution of optimal SCAs in the project instances? 
\end{itemize}

\textbf{RQ3: SCA recommendation model construction.}
\begin{itemize}
    \item \textbf{RQ3.1}: How does the SCA recommendation model perform? 
    \item \textbf{RQ3.2}: How do different ML models affect the SCA recommendation performance?
    \item \textbf{RQ3.3}: How does $\beta$ in the SCA effectiveness evaluation affect the SCA recommendation performance?
\end{itemize}

% are corresponding to the overview of our approach in Figure \ref{overview}, which
% the value distribution of the most representative features. RQ2.2 further enlighten the feature distribution between optimal SCAs and corresponding projects.
RQ1$\sim$RQ3 focus on the SCA effectiveness evaluation, SCA preference mining, and SCA recommendation model construction respectively. 
Specifically, RQ1 empirically evaluates the effectiveness of SCAs via $F_{\beta}$.
In RQ2, RQ2.1 shows the specific features selected by RFE. RQ2.2 shows the feature distribution, thereby enlightening the relation between projects under test and the correspondingly optimal SCAs.
In RQ3, by comparing four typical baselines, RQ3.1 evaluates the effectiveness of our constructed SCA recommendation model. 
RQ3.2 investigates whether different ML models affect our constructed SCA recommendation effectiveness. 
RQ3.3 investigates the impact of $\beta$ on our constructed SCA recommendation effectiveness.

\subsection{Baselines} \label{baseline} 
To the best of our knowledge, there have been currently no studies that are related to the SCA recommendation. 
Inspired by the baseline selection in the test prioritization technique recommendation community \citep{chenjunjie}, we consider four typical baselines to evaluate the effectiveness of our constructed SCA recommendation model.
Specifically, we separately regard SpotBugs, PMD, and SonarQube as the recommended SCA for a specific project. Thus, there are three typical baselines. 
In addition, we randomly select one of three candidate SCAs as the recommended SCA for a specific project. We call this baseline ``Random''.

\subsection{ML models} \label{model}
To perform the SCA recommendation, we attempt five ML models for the multi-label classification, including the tree-based Decision Tree (DT) \citep{dt}, the instance-based K-Nearest Neighbor (KNN) \citep{knn}, the linear analysis-based Logistic Regression (LR) \citep{lr}, the neural network-based Multi-Layer Perceptron neural network (MLP) \citep{mlp}, and the ensemble learning-based Random Forest (RF) \citep{rf}.
% , and the classification hyperplane-based Support Vector Machine (SVM) \citep{svm}.
\begin{table*}[]
    \centering
    \caption{The warning distribution across 213 projects. All, Min, Max, Avg, and Med are the all, minimum, maximum, average, and median values of the number of warnings.}
    \label{tab:dataset}
      \scalebox{0.75}{
    \setlength{\tabcolsep}{0.4mm}{
    \begin{tabular}{|c|c|c|c|c|c|c|c|c|c|c|c|}
    \hline
         \multirow{2}{*}{\textbf{SCA name}}	&	\multirow{2}{*}{\textbf{No. of all warnings}}	&	\multicolumn{5}{c|}{\textbf{No. of actionable warnings}}		&	\multicolumn{5}{c|}{\textbf{No. of unactionable warnings}}								\\	\cline{3-12}
			& &	\textbf{All}	&	\textbf{Min}	&	\textbf{Max}	&	\textbf{Avg}	&	\textbf{Med}	&	\textbf{All}	&	\textbf{Min}	&	\textbf{Max}	&	\textbf{Avg}	&	\textbf{Med}	\\	\hline \hline
SpotBugs	&	95273	&	6012	&	0	&	1060	&	44	&	7	&	89261	&	0	&	12769	&	449	&	173	\\	\hline									
PMD	&	316633	&	7509	&	0	&	1028	&	54	&	12	&	309124	&	0	&	26580	&	1585	&	662	\\	\hline									
SonarQube	&	132689	&	6733	&	0	&	1193	&	54	&	16	&	125956	&	0	&	8311	&	787	&	366	\\	\hline	\hline
All	&	544595	&	20254	&	0	&	3281	&	152	&	35	&	524341	&	0	&	47660	&	2821	&	1201	\\	\hline									
    \end{tabular}
    }
    }
\end{table*}

\begin{itemize}
     \item \textbf{DT} is a non-parametric supervised learning model. The goal of DT is to create a model by learning decision rules inferred from the project characteristics and use this model for the SCA recommendation.

    \item \textbf{KNN} is a simple and straightforward non-parametric supervised learning model. which uses the proximity to classify an individual sample. 
    That is, the label of a targeted sample is determined based on the labels of the nearest several labeled samples. 

    \item \textbf{LR} is a linear regression model, which is used to describe the relationship between one dependent binary variable (i.e., the optimal SCA) and one or multiple independent variables (i.e., the project characteristics).

    \item \textbf{MLP} is a type of neural network that consists of an input layer, one or more hidden layers, and an output layer. Each neuron in a layer is connected to every neuron in the next layer, forming a fully connected network.

    \item \textbf{RF} is an ensemble learning model, where the forest in RF is trained via bagging. The idea behind RF is to combine multiple weak learners to build a strong learner. 
    
    % \item \textbf{NB} is a parametric supervised learning model based on the bayes theorem with naive independence assumptions among features of project instances. 
    % In essence, NB is a conditional probability-based ML model.

    % a regression analysis model when the dependent variable is dichotomous. 

    % \item \textbf{SVM} is to find a hyperplane that best separates samples into classes with the largest margin. 
    % This hyperplane is identified in a high-dimensional space, which may be the original feature space or a higher dimensional space obtained via the transformation of a kernel function.
\end{itemize}

\subsection{Dataset and Features} \label{data} 
\begin{figure}
    % \centering
    % \includegraphics[width=0.6\linewidth]{fig/rq1-venn.png}
    % \caption{The venn graph of warnings reported by SpotBugs, PMD, and SonarQube across 213 projects.}
    % \label{fig:rq1-venn}
    \centering
    \subfloat[Actionable warnings.]{
    \label{fig:vennactionable}
    \includegraphics[width=0.4\linewidth]{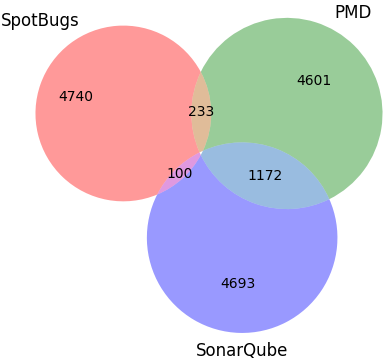}}
    \hfill
    \subfloat[Unactionable warnings.]{
    \label{fig:method4complex}
    \includegraphics[width=0.4\linewidth]{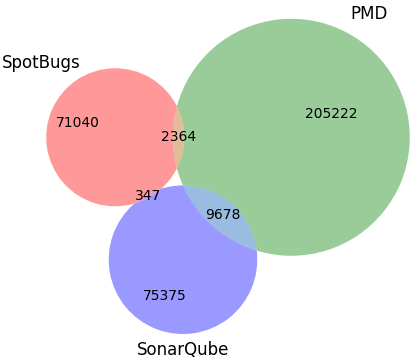}}
    \caption{The venn graph of distinct warnings reported by SpotBugs, PMD, and SonarQube across 213 projects.}
    \label{fig:venn}
\end{figure}

The SCA effectiveness evaluation in Section \ref{evaluation} performs the data preparation. 
Specifically, as shown in Section \ref{projectselection}, our approach selects 213 projects. 
The details about 213 projects can be seen in the GitHub repository \citep{mylink}. 
As shown in Section \ref{scaselection}, our approach selects three typical SCAs, including SpotBugs, PMD, and SonarQube. 
% 画个箱图，展示有效警告数在213项目的划分。
By performing the warning labeling of Section \ref{labeling}, our approach collects 544595 warnings from 213 projects across three SCAs, including 20254 actionable ones and 524341 unactionable ones. 
Table \ref{tab:dataset} shows the detailed warning distribution across 213 projects. 
After the warning alignment in Section \ref{alignment}, our approach collects 379565 distinct warnings, including 14467 actionable ones and 365098 unactionable ones.
The details of three SCAs across 213 projects are shown in Figure \ref{fig:venn}.
Based on the effectiveness evaluation in Section \ref{evaluation}, our approach identifies the correspondingly optimal SCA for each project under test via \emph{F$_{\beta}$}. Since there are 213 projects, our approach collects a total of 213 samples as the evaluation dataset.
% There are 213 projects. 

The SCA preference mining in Section \ref{mining} performs the feature preparation. 
Specifically, based on the feature extraction of Section \ref{extract}, our approach collects 302 project characteristics as features. 
By performing the feature selection of Section \ref{select}, our approach identifies 100 most representative features. 
As shown in the feature visualization of Section \ref{visual}, our approach further depicts and interprets the relation between optimal SCAs and corresponding projects.

To the end, the collected 213 samples in Section \ref{evaluation} and the selected 100 features in Section \ref{mining} are used to support the SCA recommendation model construction in Section \ref{construction}.

\subsection{Evaluation Metrics} \label{em}
We adopt \emph{P$_{micro}$}, \emph{R$_{micro}$}, and \emph{F1$_{micro}$} to evaluate the performance of SCA recommendation model \citep{takahashi2022confidence}. 
\emph{P$_{micro}$}, \emph{R$_{micro}$}, and \emph{F1$_{micro}$} are shown in Equations (\ref{mul-p}), (\ref{mul-r}), and (\ref{mul-f1}). 
Of these, \emph{k} is the number of classes. In our approach, \emph{k} is 3. 
The greater values in these three metrics indicate a more effective SCA recommendation model.

% True Positive (\emph{TP}) True Negative (\emph{TN})  False Negative (\emph{FN}) False Positive (\emph{FP})
% To describe these three metrics, we define four basic terms. True Positive (\emph{TP}) and False Negative (\emph{FN}) describe the number of samples that are correctly and incorrectly predicted by the SCA recommendation model respectively. True Negative (\emph{TN}) and False Positive (\emph{FP}) describe the number of samples that are correctly and incorrectly predicted by he SCA recommendation model respectively. 

\begin{equation}
    P_{micro} = \frac{ \sum_{k \in n} TP_k }  { \sum_{k \in n} TP_k + \sum_{k \in n} FP_k }
    \label{mul-p}
\end{equation}

\begin{equation}
    R_{micro} = \frac{ \sum_{k \in n} TP_k }  { \sum_{k \in n} TP_k + \sum_{k \in n} FN_k }
    \label{mul-r}
\end{equation}

\begin{equation}
    F1_{micro} = \frac{ 2*P_{micro}*R_{micro} }{ P_{micro}+R_{micro} }
    \label{mul-f1}
\end{equation}

\subsection{Implementation Details} \label{ed}  
\emph{Apache Maven} is used to identify the compilable release for each project. 
Our approach uses SpotBugs with version 4.0, PMD with version 7.1.0, and SonarQube with version 7.8.0 in the default settings. 
RF is embedded with the feature selection in the RFE. 
All RQs except RQ3.3, $\beta$ is set to 1, which indicates that \emph{P} and \emph{R} in \emph{$F_{\beta}$} have the same weight. 
In particular, to mitigate the accidental errors in RQ3.1, the ``Random'' baseline is repeated 100 times and the average value is regarded as the final result. 
Our approach performs 10-fold cross validation and the average value is calculated for the experimental evaluation.
As for the feature visualization and ML techniques, our approach implements them via the scikit-learn library\footnote{https://scikit-learn.org/stable/.} with the default configurations. 
All experiments are conducted on Windows 64 operating system with 12 cores (3.10GHz Gen Intel(R) CPU) and 16GB RAM.

\section{Results and analysis} \label{result}
\subsection{RQ1: SCA Performance Evaluation} \label{rq1}
% 每个工具又多少个有效警告和无效警告;归一化的警告互相之间有多少是相同---实验结果章节
As shown in Table \ref{tab:rq1}, different SCAs show varying effectiveness across 213 projects. 
Specifically, SonarQube achieves the best average F$_{\beta}$ across 213 projects. 
In particular, based on the maximum of F$_{\beta}$, SonarQube detects all defects (i.e., the union of actionable warnings from three SCAs) of certain projects, which is reflected on a project called ``gooddata''. 
In addition, it is observed that SonarQube achieves the best \emph{F$_{\beta}$} on 123 projects, followed by SpotBugs and PMD. 
It indicates that multiple SCAs obtain the same F$_{\beta}$ in 88 projects. 
Of these, the three SCAs obtain the same F$_{\beta}$ in 44 projects, while two of three SCAs obtain the same F$_{\beta}$ in the remaining 44 projects. 
Such a phenomenon is caused by the fact that \emph{P} or \emph{R} of some SCAs in the projects under test is 0. 
Overall, SonarQube performs the best among three SCAs. 
% venn graph---need to complete 

\begin{table}[]
    \caption{The effectiveness evaluation results of SCAs across 213 projects. Min, Max, Avg, and Med are the minimum, maximum, average, and median values of \emph{F$_{\beta}$}. No. of times is the number of times that a SCA achieves the best \emph{F$_{\beta}$} on 213 projects.}
    \label{tab:rq1}
    \centering
      \scalebox{0.75}{
    \begin{tabular}{|c|c|c|c|c|c|}
    \hline
         \multirow{2}{*}{\textbf{SCA name}} 	&	\multicolumn{4}{c|}{\textbf{F$_{\beta}$ (${\beta}$=1)}}						&	\multirow{2}{*}{\textbf{No. of times}}  \\ \cline{2-5}	
	& 	\textbf{Min}	&	\textbf{Max}	&	\textbf{Avg}		& \textbf{Med} & \\	\hline \hline
SpotBugs	&	0.00 	&	0.96 	&	0.07 	&	0.02 	&	104	\\ \hline
PMD	&	0.00 	&	0.82 	&	0.06 	&	0.01 	&	74	\\ \hline
SonarQube	&	0.00	&	1.00 	&	0.08 	&	0.02 	&	123	\\ \hline
    \end{tabular}
    }
    % }
\end{table}

\begin{tcolorbox}[colback=gray!13, colframe=black, boxrule=0.3mm, boxsep= -0.1cm, middle=-0.1cm]
\textbf{Summary for RQ1}: Different SCAs have varying effectiveness. Compared to SpotBugs and PMD, SonarQube can achieve superior effectiveness in terms of average $F_{\beta}$.
\end{tcolorbox}

\subsection{RQ2: SCA Preference Mining} \label{rq2}

\begin{table*}[]
    \centering
    \caption{Names and descriptions in the nine most representative features.}
    \label{tab:rq21}
      \scalebox{0.7}{
    \setlength{\tabcolsep}{0.6mm}{
    \begin{tabular}{|c|c|p{200pt}|}
    \hline
        \textbf{No.} & \textbf{Features} & \textbf{Descriptions} \\ \hline \hline
      1 & Class\_CountLineCodeDecl\_average & The average number of lines containing declarative source code in the class level\\ \hline
         2&Class\_CountDeclClassVariable\_average & Th e average number of class variables in the class level \\ \hline
       3 &Class\_CountDeclInstanceMethod\_average & The average number of instance methods in the class level\\ \hline
    4 &Class\_CountDeclInstanceVariable\_average & The average number of instance variables in the class level\\ \hline
      5 &Class\_CountDeclMethodAll\_average & The average number of methods (including inherited ones) in the class level\\ \hline
     6 & Class\_CountDeclMethodDefault\_average & The average number of local default methods in the class level\\ \hline
      7 &Class\_CountDeclMethodPrivate\_average & The average number of local (not inherited) private methods in the class level\\ \hline
      8 &Class\_CountDeclMethodProtected\_average & The average number of local protected methods in the class level\\ \hline
     9 &Class\_CountDeclMethodPublic\_average &  The average number of local (not inherited) public methods in the class level \\ \hline
    \end{tabular}
    }
    }
\end{table*}

\subsubsection{Answering RQ2.1} 
\begin{figure}
    \centering
    \includegraphics[width=0.5\linewidth]{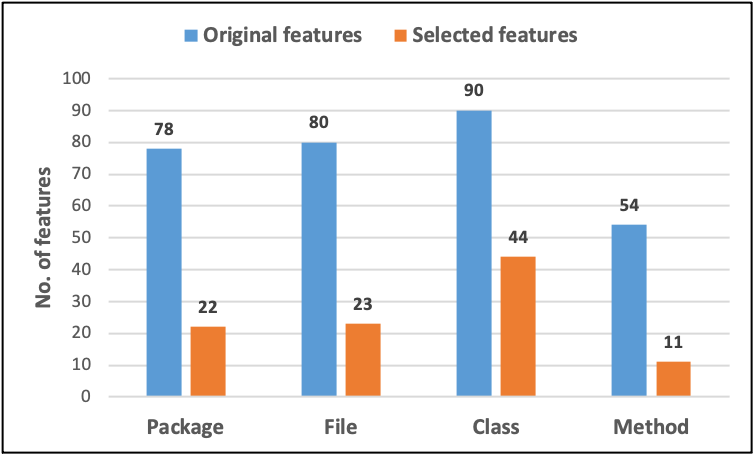}
    \caption{The number of original and selected features.}
    \label{fig:rq21}
\end{figure}

After the feature selection in Section \ref{select}, our approach identifies 100 representative features from 302 original features. 
As shown in Figure \ref{fig:rq21}, these 100 features involve 22 package-related, 23 file-related, 44 class-related, and 11 method-related software metrics.
Table \ref{rq2} shows the nine most representative features, of which all are class-related project characteristics.
It indicates that the class-related project characteristics play the most important role in the SCA recommendation.

\begin{tcolorbox}[colback=gray!13, colframe=black, boxrule=0.3mm, boxsep= -0.1cm, middle=-0.1cm]
\textbf{Summary for RQ2.1}: Our approach selects 100 features for the SCA recommendation from 302 original features. Of these, the nine most representative features fall into the class-related project characteristics.
\end{tcolorbox}

\begin{figure*}
    \centering
    \includegraphics[width=1\linewidth]{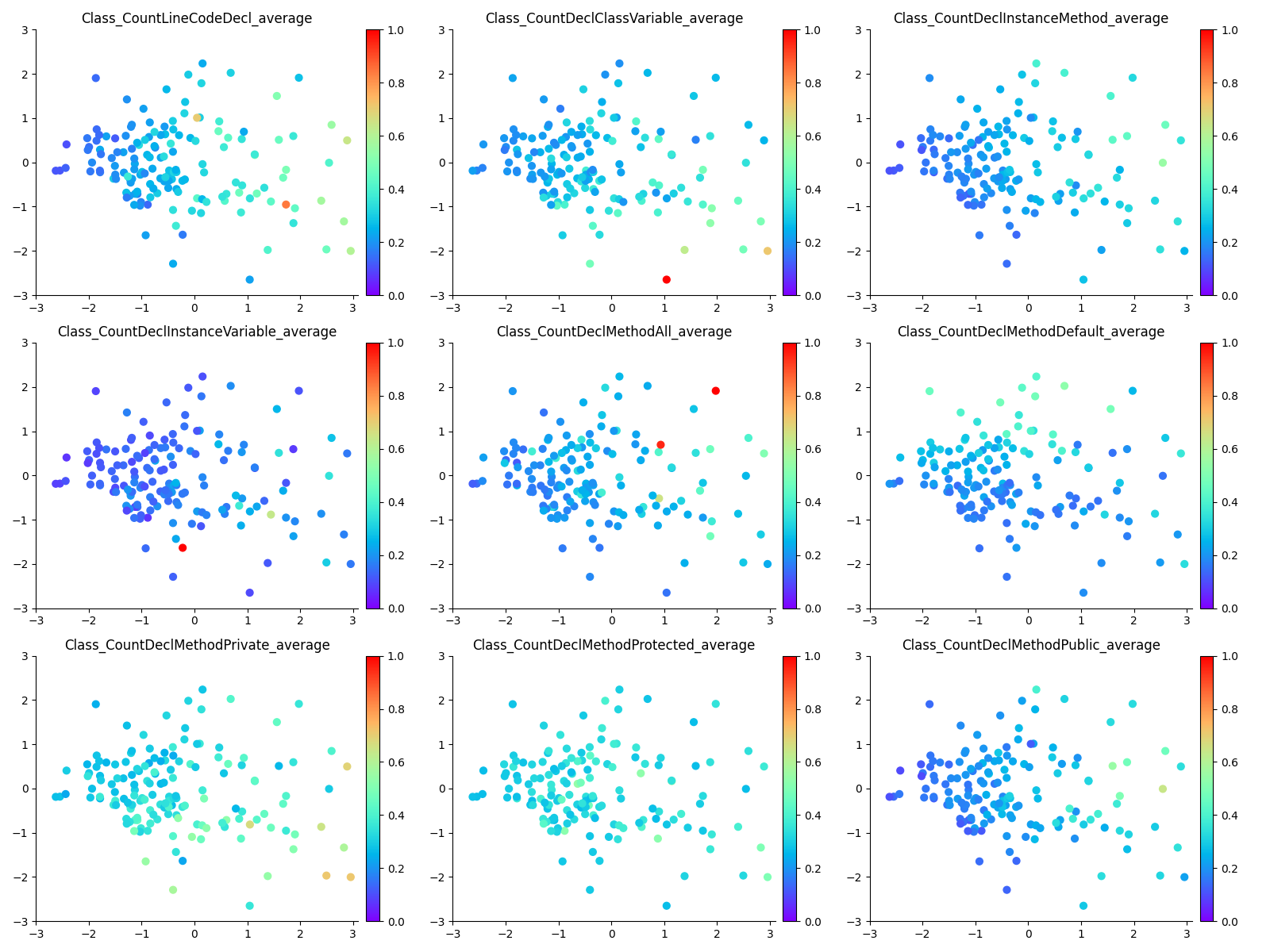}    
    \caption{The distribution of nine most representative features on 213 projects. The values of features, which are shown in the color scheme, are normalised between 0 and 1.}
    \label{fig:rq22-feature}
\end{figure*}

\begin{figure*}
    \centering
    \includegraphics[width=1\linewidth]{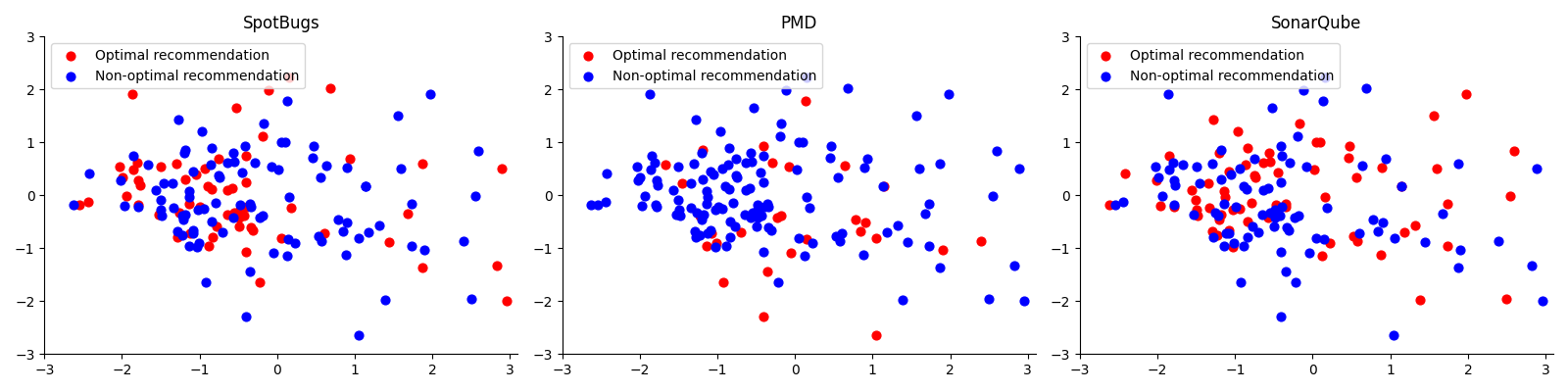}
    \caption{The distribution of the optimal SCAs on 213 projects.}
    \label{fig:rq22-sca}
\end{figure*}

\subsubsection{Answering RQ2.2} 
In the selected 100 features, our approach conducts the case study on the nine most representative features. 
Figure \ref{fig:rq22-feature} and Figure \ref{fig:rq22-sca} visualize the value distribution of the nine most representative features and the distribution of optimal SCAs on 213 projects respectively. 
Of these, each circle in the two figures denotes a project. 
As shown in Figure \ref{fig:rq22-feature}, the first, second, third, fifth, and ninth features in Table \ref{tab:rq21} have a similar value distribution. 
Also, the seventh and eighth features are similar. 
However, the value distribution of the fourth and sixth features differs from that of the other features. 
In Figure \ref{fig:rq22-sca}, a red circle represents that a current SCA achieves the optimal performance on a project under test.
Specifically, SpotBugs and SonarQube have a similar distribution, where SonarQube has slightly more red circles than SpotBugs.
However, the distribution of PMD is different from that of SpotBugs and SonarQube. 
In addition, some SCAs simultaneously achieve optimal performance on the same project.
The above phenomena correspond with the results in Table \ref{tab:rq1}. 
By combining Figure \ref{fig:rq22-feature} and Figure \ref{fig:rq22-sca}, it can be observed that when the nine features have a greater value, SonarQube is more likely to be optimal SCAs for a given project. 
For example, there is a light red circle in the distribution of ``Class\_CountLineCodeDecl\_average''. It indicates that the project with such a light red circle has a value close to 1. 
Correspondingly, SonarQube is recommended for this project. 
The above phenomenon is also reflected in the other features (e.g., Class\_CountDeclMethodAll\_average).
In contrast, PMD performs badly when ``Class\_CountDeclInstanceVariable\_average'' has a low value. It indicates that when a project under test has a low average number of instance variables at the class level, it is more difficult for PMD to detect defects in this project than SonarQube.
The above results reveal that the effectiveness of SCAs is impacted by features and no SCAs can be considered the best in all projects.
That is, SCAs have different preferences in projects under test, which makes it possible to select the most suitable SCA for a project with particular characteristics.

% That is, the optimal SCAs determined in a project are dependable of project characteristics. 
% Besides, 
% it is observed that SCAs 
% different strengths and weaknesses of existing APRTs, which calls for
% methods that make it possible to select the most suitable technique given a buggy program
% with particular features.
% Thus, it is  make it possible to select the most suitable technique given a buggy program with particular features.

\begin{tcolorbox}[colback=gray!13, colframe=black, boxrule=0.3mm, boxsep= -0.1cm, middle=-0.1cm]
\textbf{Summary for RQ2.2}: The feature distribution is different, which is closely related to the optimal SCA determination on a given project with particular characteristics.
\end{tcolorbox}

\subsection{RQ3: SCA Recommendation Model Construction} \label{rq3} 

\begin{figure}
    \centering
    \includegraphics[width=0.5\linewidth]{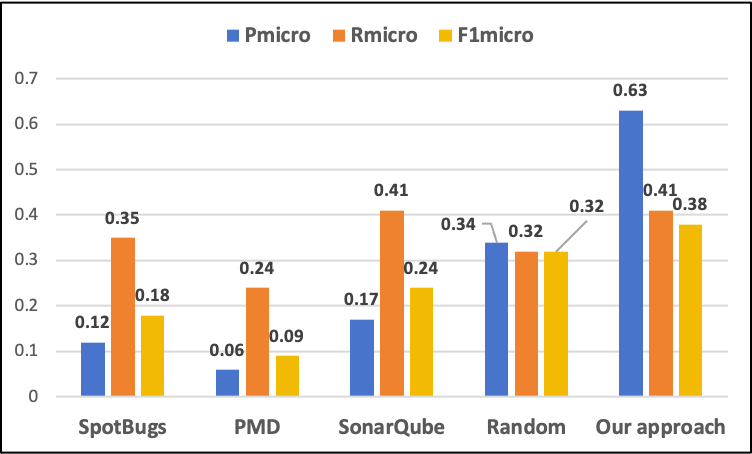}
    \caption{The SCA recommendation results in the four baselines and our approach.}
    \label{fig:rq31}
\end{figure}

\subsubsection{Answering RQ3.1}
Figure \ref{fig:rq31} shows the SCA recommendation results between four typical baselines and our approach. 
Of these, the ``Random'' baseline is repeated 100 times and the average result is calculated as the final result. 
In the remaining three baselines, Table \ref{tab:rq1} and Figure \ref{fig:rq22-sca} show that different SCAs may show the same F$_{\beta}$ on the same project. 
To alleviate the above problem, the following operations are considered. 
For example, if SpotBugs and SonarQube simultaneously achieve the same F$_{\beta}$, we separately consider SpotBugs and SonarQube as the optimal result for this project when SpotBugs and SonarQube are used as baselines respectively. 
As shown in Figure \ref{fig:rq31}, our approach consistently performs better than the four baselines. 
Specifically, our approach achieves 0.63 of P$_{micro}$, which exceeds SpotBugs, PMD, and SonarQube by about 5 times, 11 times, and 4 times respectively.
The R$_{micro}$ of our approach is 17\% and 71\% better than that of SpotBugs and PMD respectively. 
However, our approach has the same R$_{micro}$ with SonarQube.
In terms of F1$_{micro}$, our approach is about 2 times, 4 times, and 2 times better than SpotBugs, PMD, and SonarQube respectively.
Also, our approach is 85\%, 28\%, 19\% better than the ``Random'' baseline in term of P$_{micro}$, R$_{micro}$, and F1$_{micro}$ respectively.
Thus, our approach can achieve the effective SCA recommendation for projects under test.

\begin{tcolorbox}[colback=gray!13, colframe=black, boxrule=0.3mm, boxsep= -0.1cm, middle=-0.1cm]
\textbf{Summary for RQ3.1}: Our approach achieves 0.63 of P$_{micro}$, 0.41 of R$_{micro}$, and 0.38 of F1$_{micro}$, which consistently performs more effective SCA recommendation than the four typical baselines.
\end{tcolorbox}

\begin{figure}
    \centering
    \includegraphics[width=0.5\linewidth]{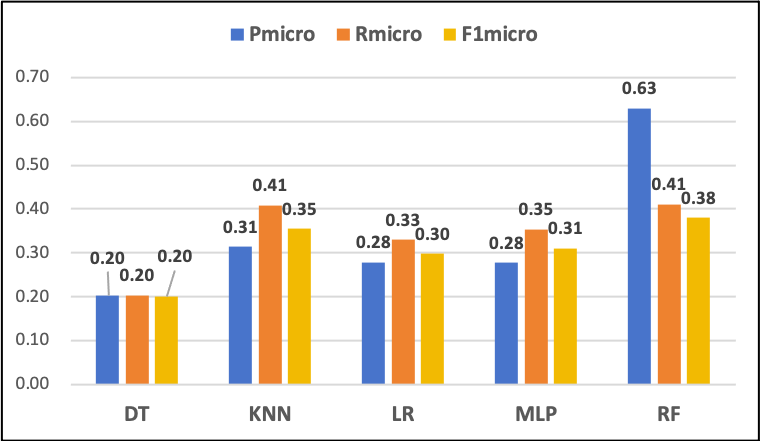}
    \caption{SCA recommendation model performance under five ML models.}
    \label{fig:rq32}
\end{figure}
\subsubsection{Answering RQ3.2} Figure \ref{fig:rq32} shows the performance of SCA recommendation model under different ML models. 
Specifically, RF achieves P$_{micro}$ with 0.63, which exceeds the other four ML models by 2 $\sim$ 3 times.
R$_{micro}$ of RF is 0.41, which is the same as R$_{micro}$ of KNN. However, R$_{micro}$ of RF is 100\%, 24\%, and 17\% better than that of DT, LR, and MLP respectively. 
RF obtains F1$_{micro}$ with 0.38, which performs better than the other four ML models by 9\% $\sim$ 90\%.
The above results indicate that compared to the other four ML models, RF can achieve superior performance in the SCA recommendation.

\begin{tcolorbox}[colback=gray!13, colframe=black, boxrule=0.3mm, boxsep= -0.1cm, middle=-0.1cm]
\textbf{Summary for RQ3.2}: In the five ML models, RF with 0.63 P$_{micro}$ can achieve the optimal SCA recommendation performance.
\end{tcolorbox}

\subsubsection{Answering RQ3.3}
\begin{figure}
    \centering
    \includegraphics[width=0.5\linewidth]{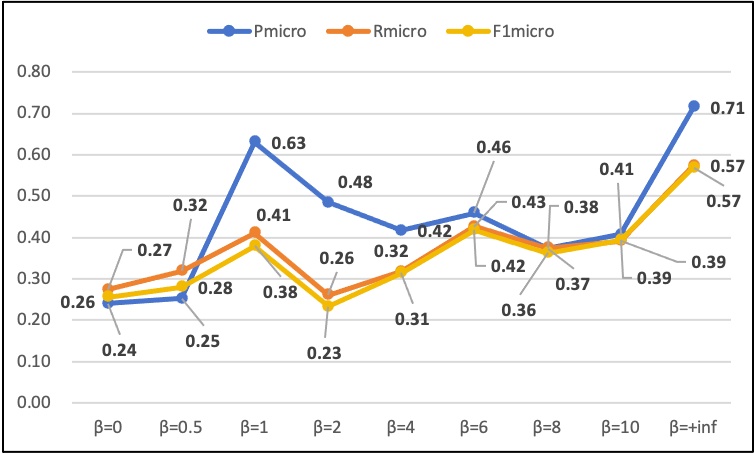}
    \caption{Performance of SCA recommendation model under different $\beta$ of F$_{\beta}$.}
    \label{fig:rq33}
\end{figure}

% features are the same
To answer RQ3.3, our approach investigates the impact of $\beta$ on the performance of SCA recommendation model.
As shown in RQ 3.2, RF performs better than the other four ML models. Thus, this RQ is conducted on RF.
Figure \ref{fig:rq33} shows the performance of SCA recommendation model under different $\beta$ of F$_{\beta}$. Particularly, \emph{$\beta$=0} indicates that our approach uses \emph{P} to evaluate the effectiveness of SCAs.
\emph{$\beta$=1} indicates that \emph{P} and \emph{R} are considered to be the same in the SCA effectiveness evaluation. 
\emph{$\beta$=+inf} indicates that our approach uses \emph{R} to evaluation the effectiveness of SCAs. 
Specifically, when \emph{$\beta$} varies from 0 to \emph{+inf}, the trend of \emph{$R_{micro}$} is very similar to that of \emph{$F1_{micro}$}. 
In contrast, the trend of \emph{$P_{micro}$} is different from that of \emph{$R_{micro}$} and \emph{$F1_{micro}$}. 
It is noted that compared to other values of \emph{$\beta$}, the performance of SCA recommendation model is the best when \emph{$\beta$=+inf}. 
When \emph{$\beta$} is from \emph{0} to \emph{+inf}, the fluctuation range of {$P_{micro}$} is 0.26 $\sim$0.71, while {$R_{micro}$} and \emph{$F_{micro}$} fluctuate between 0.24 and 0.57.
It indicates that $\beta$ significantly affects \emph{$F_{micro}$}, while having a slight impact on \emph{$P_{micro}$} and \emph{$R_{micro}$}. 

\begin{tcolorbox}[colback=gray!13, colframe=black, boxrule=0.3mm, boxsep= -0.1cm, middle=-0.1cm]
\textbf{Summary for RQ3.3}: $\beta$ has a significant impact on the SCA recommendation model performance. In particular, when \emph{$\beta$=+inf}, the SCA recommendation model can obtain optimal performance.
\end{tcolorbox}

\section{Discussion} \label{discussion}

\subsection{Limitations and improvement directions} 
Our approach aims to perform a practical SCA recommendation for a given project. Besides, the above results in Section \ref{result} show that our approach performs better than the four typical baselines.
However, our approach may still face some limitations in terms of SCA performance evaluation and SCA preference mining. Based on these limitations, we further provide the improvement directions to enhance the SCA recommendation.

% \footnote{https://www.nist.gov/system/files/documents/2021/03/24/CAS\%202012\%20Static\%20Analysis\%20Tool\%20Study\%20Methodology.pdf}
On the one hand, our approach determines the optimal SCA for a given project by evaluating the effectiveness of all candidate SCAs on this project.
The SCA effectiveness is the defect detection ability of SCAs, which is the most commonly used to evaluate SCAs.
In addition to the SCA effectiveness, some other aspects (e.g., SCA efficiency, SCA discrimination, SCA overlap, and SCA coverage) are considered to evaluate SCAs \citep{7181477}.
The SCA efficiency measures the time overhead of defect detection for a given project. 
The SCA discrimination reflects whether an SCA can detect defects when these defects exist but remain silent when a similar source code is used safely.
The SCA overlap is the proportion of defects detected by more than one SCA, which can help reveal the statistical independence of SCAs.
The SCA coverage measures the types of defects detected by a SCA.
In a project with high complexity, SCAs with more delicate techniques (e.g., flow sensitivity) generally perform more effectively than SCAs with simple techniques (e.g., pattern matching) \citep{likaixuan}.
In turn, SCAs with more delicate techniques undertake more time overhead than SCAs with simple techniques. 
Also, SCAs with different techniques could bring inconsistent discrimination, overlap, and coverage \citep{7181477}.
As such, more evaluation metrics should be considered to comprehensively evaluate the performance of SCAs. 
Further, it could be useful to design a new metric for fusing the current SCA evaluation metrics, thereby understanding the holistic performance of SCAs.

On the other hand, our approach focuses on extracting project characteristics to mine the intrinsic relation between projects under test and the correspondingly optimal SCAs. 
It indicates that the current SCA recommendation in our approach does not take specific project scenarios into account. 
In practice, different projects have different scenarios, including the quality assurance requirements and available resources \citep{Nunes2018BenchmarkingSA}. 
For example, some projects are required high-quality, and SCAs with the highest defect detection rate should be selected for these projects.
Some projects have limited resources, and SCAs with the minimum time overhead should be recommended for these projects. 
Thus, it should consider the project scenario-related features and mine the preference between SCAs and projects, thereby achieving a more flexible SCA recommendation.

% 现在只是从项目的特性考虑的，现在的sca的推荐模型构建是与SCA无关的
% 现在的只考虑了进行了case study的研究，需要考虑更多的sca
%数据集是Ade基础，累计更多的更加丰富的数据集是非常有必要的
\subsection{Threats to the validity}

\textbf{External.} A threat to external validity refers to the result generalization of our approach. 
To alleviate this threat, we design rigorous criteria to select 213 open-source and large-scale projects. Besides, we carefully select three typical SCAs, which are equipped with syntax and semantics-based static analysis techniques. 
In the future, we will explore and extend the generalization of our approach in more projects and SCAs.
% As such, we believe that our approach has sufficient generalization for SCA recommendation. 

\textbf{Internal.} The threat to internal validity contains two aspects.
The first aspect is the stochastic nature of the ML model. 
To avoid this bias, we conduct 10-fold cross validation and calculate the average values of metrics as final results.
The second aspect is the parameter bias (i.e., $\beta$ of F$_{\beta}$). To mitigate this problem, we investigate the impact of different $\beta$ on the SCA recommendation model performance.

\textbf{Construct.} The threat to construct validity is related to the dataset quality. 
To collect the dataset for SCA recommendation construction, we perform the warning labeling in Section \ref{labeling} via the closed-warning heuristic \citep{dataset11}, which is the most advanced method to automatically label warnings \citep{ge2023machine}. 
Besides, during the effectiveness evaluation of SCAs in a project, we perform the warning alignment and consider all distinct actionable warnings reported by candidate SCAs to approximate all true defects of this project. 
Such an operation can gather the SCA defect detection ability to mitigate the oracle of all true defects in a project to some extent.
However, we acknowledge that there could still be a few noisy samples in the dataset. 
These noisy warnings could facilitate the robustness of the ML-based SCA recommendation model \citep{xu2021differential}.
In the future, we plan to collect more ground-truth datasets for the SCA recommendation.

\section{Conclusion} \label{conclusion}
In this paper, we propose a practical SCA recommendation approach, which aims to construct the ML-based SCA recommendation model based on the results of SCA effectiveness evaluation and SCA preference mining. 
Though conducting the experimental evaluation on 213 open-source and large-scale projects and three typical SCAs, the results show that in terms of $P_{micro}$, $R_{micro}$, $F1_{micro}$, our approach performs better than the four typical baselines.

% We would like to thank the anonymous reviewers for their insightful comments. 
\section*{Acknowledgements} The work is partly supporteid by the National Natural Science Foundation of China (61932012, 62372228, 62141215).

\bibliography{ref}

\end{document}